\documentclass[11pt,twoside]{article}
\usepackage{asp2010}

\resetcounters

\newcommand{\mdot}{\mbox{$M_{\odot}$ yr$^{-1}$}}
\newcommand{\msun}{$M_{\odot}$}
\newcommand{\degr}{$^{\mathrm{o}}$}
\def\sci{Science}             

\begin{document}

\title{The interstellar dust reservoir: SPICA's view on dust production and the interstellar medium in galaxies}
\author{F.~Kemper$^{1}$, R.~Zhao-Geisler$^{2,1}$, O.~C.~Jones$^{3,1}$ and S.~Srinivasan$^{1}$
\affil{$^{1}$Institute of Astronomy and Astrophysics, Academia Sinica, Taiwan}
\affil{$^{2}$National Taiwan Normal University, Taipei, Taiwan}
\affil{$^{3}$Jodrell Bank Centre for Astrophysics, The University of Manchester, UK}}

\begin{abstract}
  Typical galaxies emit about one third of their energy in the
  infrared. The origin of this emission reprocessed starlight absorbed
  by interstellar dust grains and reradiated as thermal emission in
  the infrared. In particularly dusty galaxies, such as starburst
  galaxies, the fraction of energy emitted in the infrared can be as
  high as 90\%.  Dust emission is found to be an excellent tracer of
  the beginning and end stages of a star's life, where dust is being
  produced by post-main-sequence stars, subsequently added to the
  interstellar dust reservoir, and eventually being consumed by star
  and planet formation.  This work reviews the current understanding
  of the size and properties of this interstellar dust reservoir, by
  using the Large Magellanic Cloud as an example, and what can be
  learned about the dust properties and star formation in galaxies
  from this dust reservoir, using SPICA, building on previous work
  performed with the \emph{Herschel} and \emph{Spitzer} Space
  Telescopes, as well as the Infrared Space Observatory.
\end{abstract}

\section{The dust reservoir in galaxies}

\subsection{Sources and sinks in the Large Magellanic Cloud}

The Large Magellanic Cloud (LMC) was imaged using \emph{Spitzer}
\citep[SAGE;][]{Meixner_06_Spitzer} and \emph{Herschel}
\citep[HERITAGE;][]{Meixner_13_Herschel} using all non-overlapping
photometric bands available on these telescopes. The combined
SAGE-HERITAGE survey, often referred to as Mega-SAGE, covers the
wavelength range from 3.6 to 500 $\mu$m. The four \emph{Spitzer}-IRAC
bands (3.6--8.0 $\mu$m) mainly show circumstellar dust emission from
stellar point sources, and extended emission due to polycyclic
aromatic hydrocarbons (PAHs), thus forming a good tracer of the
production and consumption of dust in post- and pre-main-sequence
stars, while the \emph{Spitzer}-MIPS and \emph{Herschel} images show
the dust mass in the interstellar medium (ISM). The MIPS 24 $\mu$m
band represents a transition, while dominated by extended emission due
to interstellar dust, extremely dusty point sources are clearly
detectable in this band too. It is these point sources with 24-$\mu$m
detections that dominated the dust production and consumption budget.

Building on a series of previous estimates and methods to determine
dust production rates by evolved stars,
\citet{Riebel_12_Mass} estimate a total dust production rate
of $\sim 2.1 \times 10^{-5}$ \mdot from the \emph{Spitzer}-IRAC and
MIPS data. This rate should be contrasted against the dust mass in the
interstellar medium of the LMC, $\sim 10^{6}$ \mdot, derived from the
\emph{Herschel} observations \citet{Skibba_12_Spatial}, and the dust
consumption in star formation \citep[$\sim 1.9 \times
10^{-3}$;][]{Skibba_12_Spatial}, assuming a dust-to-gas ratio of
1/200.  Taking these numbers at face value, it becomes clear that
evolved stars do not produce sufficient dust to replenish the
interstellar medium, on the residence time scale of dust in the
interstellar medium \citep[$\sim$ 2.5 Gyr;][]{Tielens_90_Towards}.
Moreover, the much higher dust consumption rate in comparison to the
dust production rate even suggests that the dust reservoir in the
interstellar medium of the LMC is currently being depleted. This would
imply that until fairly recently, the dust production rate was much
higher than at the present day, or, alternatively that we are missing
a significant source of dust production in this equation.

Three alternatives suggest themselves: {\bf i)} The dust production in
supernovae has not been considered in the estimates of dust production
by evolved stars in the LMC \citep[e.g.][]{Riebel_12_Mass},
due to low number statistics, but estimates suggest that it could be
similar to the production by Asymptotic Giant Branch stars
\citep[e.g.~][]{Whittet_03_dust}. Indeed, \citet{Dunne_03_Type} argued
that supernova remnant Cas A has produced 2-4 \msun\, of dust,
although it has since been shown that the bulk of the dust mass
detected is part of a foreground cloud \citep{Krause_04_No}. From
observations of contemporary supernovae in external galaxies,
\citet{Sugerman_06_Massive} showed that a more realistic number of
dust production may lie around $10^{-3}-10^{-2}$ \msun, although
\emph{Herschel} observations of SN 1987A show that it has already
produced 0.5 \msun in the first 25 years
\citep{Matsuura_11_Herschel}. {\bf ii)} Dust production in the
interstellar medium is efficient, and the source of interstellar dust
is not stellar. This has been explored by
\citet{Zhukovska_08_Evolution} and by
\citet{Jones_05_Dust}, and seems to be a viable
option. \citet{Zhukovska_08_Evolution} estimate that the dust
production in the interstellar medium in the Solar Neighborhood
exceeds the dust production by evolved stars after as little as 1 Gyr.
{\bf iii)} Dust production by \emph{extreme} Asymptotic Giant Branch
(AGB) stars may be overlooked. Searches for dusty evolved stars rely
on classifying point infrared (IRAC, 2MASS) point sources
\citep[e.g.][]{Boyer_11_Surveying}, which are subsequently being fitted
against a dust shell model grid to determine the integrated dust
production rate \cite[e.g.][]{Sargent_11_Mass,Srinivasan_11_mass}.
From this procedure it becomes clear that the dust production is
dominated by the reddest objects, the so-called extreme AGB
stars. Indeed, \citet{Riebel_12_Mass} found that this small
number (4\%) of extremely red sources account for 75\% of the total
dust production. This immediately raises the question how many even
dustier and redder sources are present, and how they can be
detected. With the peak of their spectral energy distributions (SEDs)
shifting to even longer wavelengths they can no longer be picked up in
the near- or mid-infrared, and far-infrared detections are
required. At these wavelengths, the point spread function (PSF)
becomes very large, and the point sources need to be bright to stand
out over the extended emission due to interstellar dust within the
beam.  At the distance of the LMC, this is generally not the case, and
\citet{Boyer_10_Cold} demonstrated that only a small number of known
evolved stars in the LMC have counterparts in the \emph{Herschel}
data.

\subsection{The composition of stellar dust}

Including spectroscopy in the analysis of evolved stars enables us to
dissect the dust production not only by type of AGB star (carbon-rich
or oxygen-rich), but to narrow down the mineralogical components in
the freshly produced dust. Although a huge amount of variation exists,
and still needs to be further explored, studies like those done by
\citet{Sargent_10_Mass} and \citet{Srinivasan_10_mass}
modeled representative oxygen-rich and carbon-rich AGB star spectra
for the LMC, and determined typical compositions. Combining this with
the dust production rate derived from model grid fitting as described
above, allows for a determination of the total dust production split
out by mineral. It is found that in the LMC, of the dust produced by
AGB stars and Red Supergiants, 77 wt.\% is in the form of amorphous
carbon, 11\% in the form of silicon carbide (SiC), 12\% in amorphous
silicates, and a small fraction of $<1$\% of the freshly produced dust
may be in the form of crystalline silicates
\citep{Kemper_13_Stellar}.

Although most extreme AGB stars turn out to be carbon-rich, the exact
composition of dust produced by these stars is hard to determine due
to the high optical depth in the circumstellar dust shells. The
composition of only the optically thin outer layer can be optimally
probed (Speck et al.~\emph{in prep.}), so the SiC/amorphous carbon
ratio in the vast majority of the dust mass produced by extreme AGB
stars remains unknown, although one may assume that it remains
constant throughout the mass losing phase of the star.

As stated before, supernovae may be important dust sources too, but so
far little is known about the composition of the dust in these
environments.  One of the few studies that have looked into this is
done by \citet{Rho_08_Freshly}, who derived a
multi-component composition dominated by amorphous silicates for Cas A in
an attempt to fit the main spectral feature at 21 $\mu$m.

The interstellar dust of our own Milky Way shows a different
composition from what is produced by evolved stars; the majority of
the dust is in the form of amorphous silicates, and a smaller fraction
in amorphous carbon \citep{Tielens_05_Origin}. Only trace amounts of
SiC \citep[$\sim 2.6 - 4.2$\%;][]{Min_07_shape} and crystalline
silicates \citep[$<2$\%][]{Kemper_04_Absence} can be present.  The LMC ISM is similar
in composition to the Galactic ISM, because the ultraviolet extinction
curve between the Milky Way and the LMC is virtually identical,
including the relative strength of the 2175 \AA\, bump
\citep{Fitzpatrick_86_average}, meaning similar ratios between
carbon-rich and oxygen-rich dust in both galaxies.
The discrepancy between ISM dust and stellar dust composition is
consistent with the idea that in both galaxies dust formation in
denser parts of the ISM itself may dominate the overall dust
production, however, there is a contribution from evolved stars that
needs to be considered.

\section{Observing circumstellar and interstellar dust with SPICA}

\subsection{The dust reservoir in Local Group galaxies as 
observed by SPICA}

The Mid-Infrared Camera and Spectrograph (MCS) on board of SPICA will
operate at 5-38 $\mu$m, and this wavelength range will be extended
to the near-infrared with the Focal Plane Camera (FPC) which covers
the 0.7-5 $\mu$m range, providing continuous band coverage from 0.7-38
$\mu$m at a spectral resolution $R\approx 5-10$. The long wavelength
instrument SAFARI will operate at 34-210 $\mu$m, and has two modes
that are suitable for SED construction: first, the spectroscopic SED
mode, with $R\approx 50$, which will allow us to get SEDs from all
pixels in a map; and, second, -- the more useful option for surveys of
the dust budget in galaxies -- the photometric mode which covers the
SAFARI wavelength range in three broad bands. Thus SPICA uniquely
provides a full wavelength coverage from 0.7 -- 210 $\mu$m in
photometric bands, where for instance \emph{Spitzer} had gaps in the
wavelength coverage. In particular the gap between the IRAC-[8.0] and
MIPS-[24] band has been hampering the photometric classification of
carbon-rich and oxygen-rich AGB stars, as prominent broad spectral
features are available in this range (the 9.7 $\mu$m silicate feature,
and the 11.3 $\mu$m SiC feature). The silicate band can also be used
for further constraining the optical depth and thus the mass-loss rate
of oxygen-rich AGB stars. The photometric bands available on the suite
of instruments on board of SPICA will overcome these problems and can
take advantage of the diagnostic power of broad spectral features due
to dust, because of the continuous wavelength coverage.

With SPICA it will become possible to do studies of the interstellar
dust reservoir and dust injection by evolved stars in galaxies beyond
the Magellanic Clouds. Studying external galaxies allows for taking in a
global view of the dust reservoir and dust producing stars in the
entire galaxy, without having to account for interstellar extinction.
Obvious candidates are M31 and M33, the two
other spiral galaxies, besides the Milky Way, in the Local Group, at
distances of 752 $\pm$ 27 kpc \citep{Riess_12_Cepheid} and 840 kpc
\citep{Freedman_91_New}, respectively.  Although M31 is the
closest one of the two, it is more inclined, and it will be harder to
avoid crowding and confusion,
compared to M33.  

\begin{figure}[!ht]
\begin{center}
   \plottwo{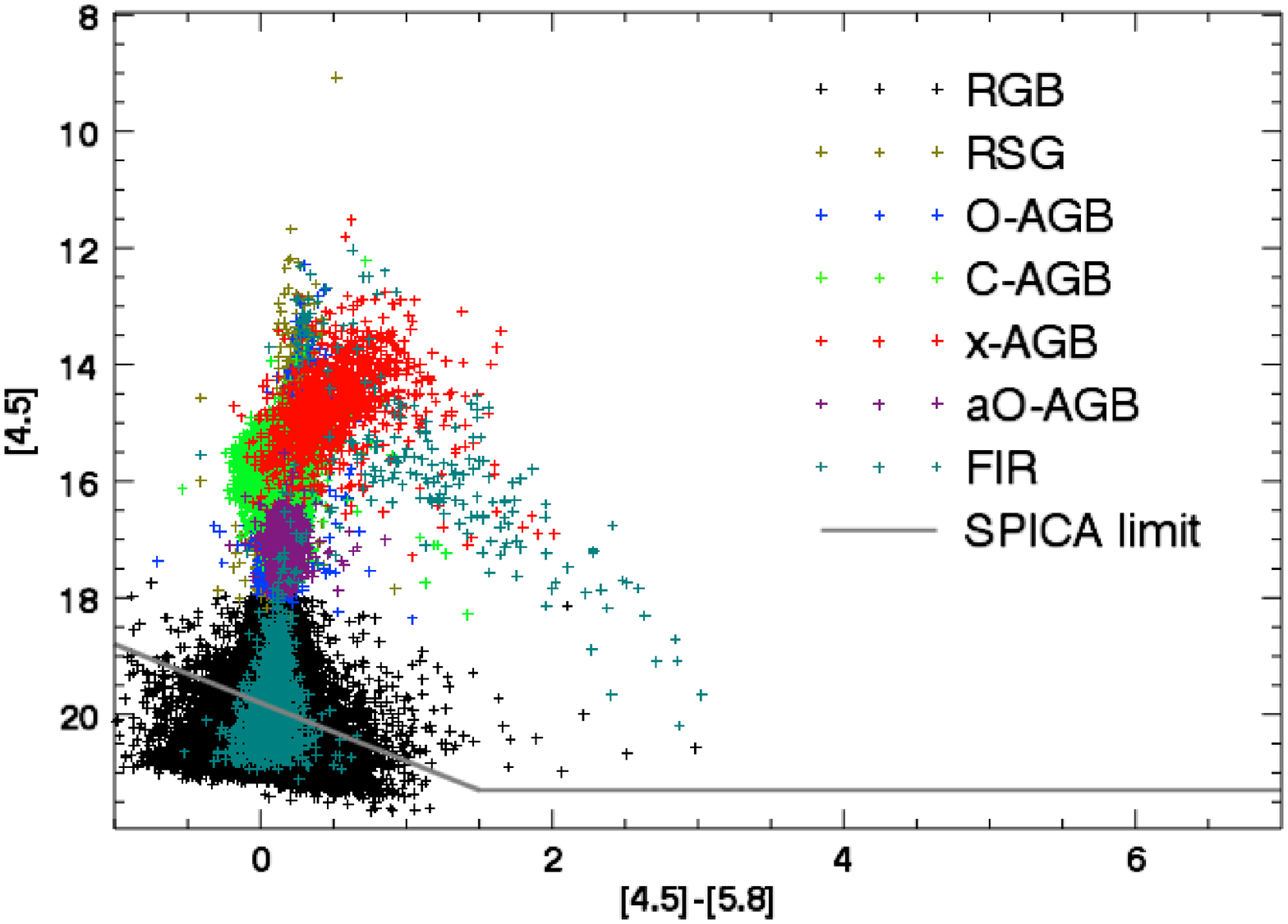}{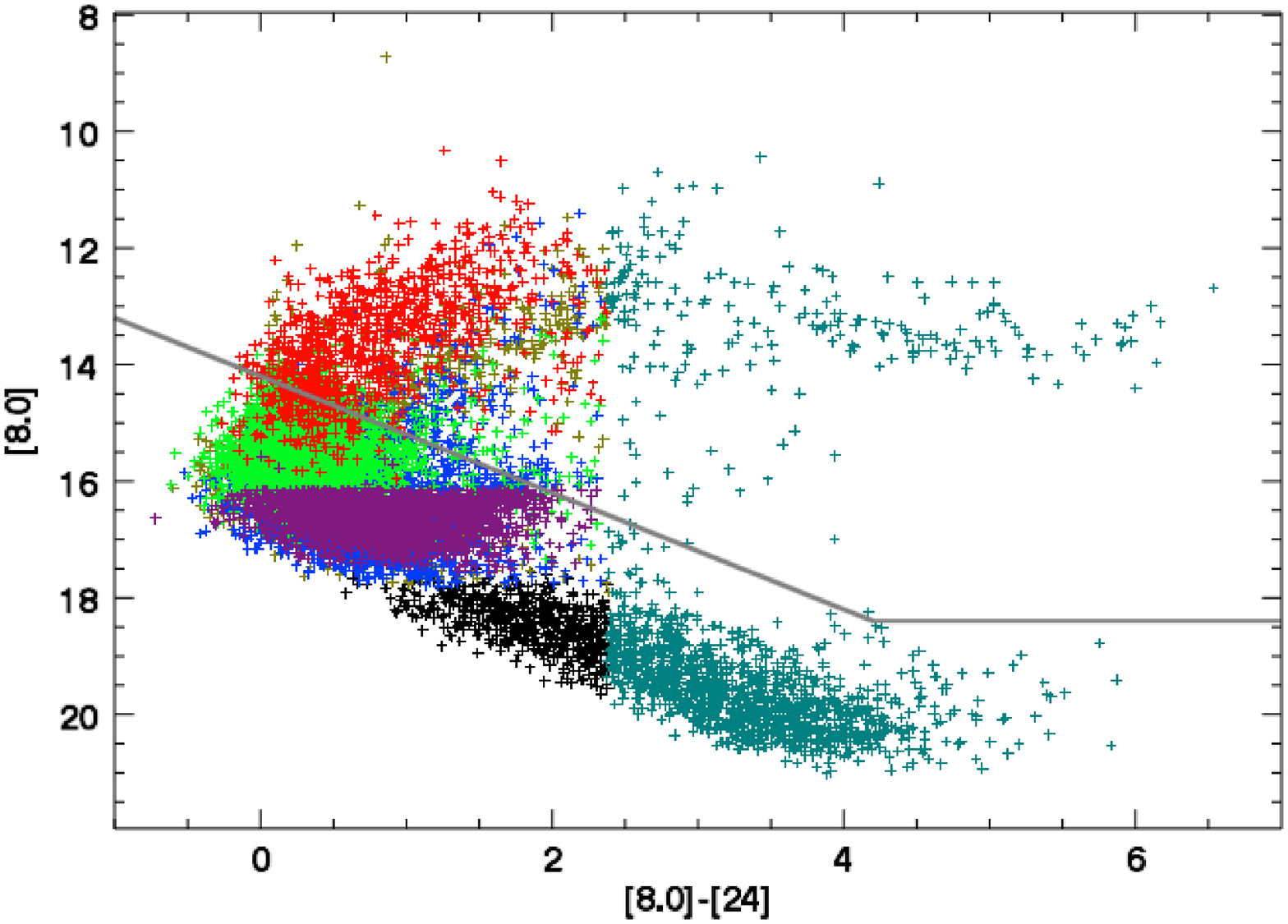}
\end{center}
\caption{Color-magnitude diagram (CMD) showing different classes of
  mass-losing evolved stars (SAGE-LMC stars at the distance of M31;
  only the stars classified by \citet{Boyer_11_Surveying} are included)
  and the discovery space of SPICA (gray line). Square brackets denote
  the brightness in Vega magnitudes at the wavelength enclosed in the
  brackets. The used \emph{Spitzer} filters match closely with the
  proposed SPICA-MCS filters. While at shorter mid-IR wavelengths the
  same stellar populations as for the SAGE LMC with Spitzer can be
  detected (left), only the stars with a very high mass-loss rate can
  be found at longer wavelengths (right).  }\label{fig:m31cmd}
\end{figure}

Thus, an exciting possibility is to set up a survey of M31 (or M33) to
make an inventory of the stellar content, particularly focused on
dusty objects, in a similar fashion as the SAGE-LMC survey
\citep{Meixner_06_Spitzer}. We have performed some initial estimates and
found that a 1\degr $\times$ 3\degr\, region encompassing the entirety
of M31 can be mapped with 600 pointings using MCS.  If we employ 4
filters (two in WFC-L, and two in WFC-S), and do observations in 2
epochs to check for spurious point sources and variability, we find
that with the minimum amount of integration time per frame we can
perform these observations in $\sim$ 400 hours. The WFC-S and WFC-L
modes can be used at the same time, as can the FPC-S instrument, with
a slight sky offset. This setup does not cover the full spectral
range, but a factor of 5 increase in total time would cover all MCS
filters, leading to a project time of $\sim 2000$ hours for the 1\degr
$\times$ 3\degr\, area covering M31 on the sky. Reductions in
integration time can be made again by deciding to cover only half or a
quarter of M31, and extrapolating the statistics on the stellar
population.

Fig.~\ref{fig:m31cmd} shows the color-magnitude diagrams (CMD) for the
point sources detected in the SAGE-LMC survey, classified by
\citet{Boyer_11_Surveying} into different types of evolved dust producing
stars. The bands presented in these diagram are within the MCS
wavelength range, and a closely matching MCS equivalent can be
identified in the currently proposed filter set. The sample has been
placed at the M31 distance and compared with the detection limits of
the proposed survey.  For the shortest integration times available on
MCS, the detection limits for the WFC-S bands are virtually the same
for the SAGE-LMC project and the M31 survey proposed in this work, in
terms of detectability of the type of object. The detection limit is
indicated by a gray line in the each of the panels of
Fig.~\ref{fig:m31cmd}.  However at longer wavelengths, it turns out
that the current detection limit for the proposed survey would not
detect the same stellar population in M31 as it did in the SAGE-LMC
survey, by at least 2 magnitudes. It appears that the choice of
filters with $R\approx 10$ are too narrow to go deep, especially in
comparison with MIPS-[24], and perhaps an additional set of 2 or 3
broad band filters for the WFC-L mode would remedy this, making the
WFC-L as much of an improvement over MIPS-[24] as that WFC-S has been
improved over IRAC-[4.5] and IRAC-[5.8].


\subsection{Spectroscopy with the Mid-infrared Camera and Spectrograph (MCS)}

The currently proposed MCS instrument includes two spectroscopic
modes: the High Resolution Spectrograph (HRS-L), which will operate
at 12-18 $\mu$m, with $R\approx 20,000 - 30,0000$. and the Medium
Resolution Spectrograph, which will operate at 12.2-23.0 $\mu$m, with
$R\approx 2000$ (MRS-S) and at 23.0-37.5 $\mu$m, with $R\approx 1000$
(MRS-L). For the purpose of studying solid state features, the
spectral resolution offered by the MRS is sufficient to do a detailed
analysis of the mineralogy. Unfortunately, the wavelength coverage of
the MRS mode does not include a number of important spectral features:

\begin{itemize}
\item The Si-O stretching mode in amorphous and crystalline silicates around
9.7 $\mu$m.
\item Features due to polycyclic aromatic hydrocarbons (PAHs) from 6-12 $\mu$m.
\item Two out of the four identified bands of C$_{60}$, namely those at 7.0
and 8.5 $\mu$m. 
\item Ice features at 4.2 $\mu$m due to CO$_2$; 6.0 $\mu$m due to H$_2$O and 6.8 $\mu$m due to a still unidentified carrier
\item Absorption lines from C$_2$H$_2$ and other molecules related to dust formation.
\item Spectral features due to alumina, oxides and quartz.
\item The 11.3 $\mu$m feature due to SiC.
\end{itemize}

The wavelength range shortward of 12.2 $\mu$m is covered by a grism in
WFC-S, with an intended spectral resolution of $R\approx 50$. While
this may, in some cases, be sufficient to detect the presence of these
features, it is not possible to do any compositional analysis, such as
for instance the crystalline fraction of silicates, which is revealed
by the shape of the resonance. For such an analysis, the observations
should match the spectral resolution of laboratory data, which is
typically taken at $R$ of a few hundred. Moreover, the detection of
narrow features, such as the C$_2$H$_2$ absorption lines, or the
C$_{60}$ narrow emission bands will be virtually impossible at $R
\approx 50$.  Thus we recommend the insertion of a $R \approx 200$
grism in the WFC-S filter wheel, with a wavelength coverage of
5.0-12.2 $\mu$m. Perhaps this may have to be divided over two grisms,
to avoid problems with the dispersion.

There are a number of other interesting dust and ice resonances that are
 included in the MRS wavelength coverage. Here we highlight a
few examples, to illustrate the discovery space that can be further
explored with the MRS mode on the MCS.

First, carbon dioxide (CO$_2$) in its pure form has a double-peaked
resonance around 15.2 $\mu$m \citep{Ehrenfreund_99_Laboratory}.  This
double-peaked structure becomes less pronounced or disappears
altogether when the a-polar CO$_2$ ice is embedded in polar
H$_2$O-rich ices. The depth of the trough between the two peaks can
provide information about the amount of thermal processing, and thus
about the degree of separation between the two ice components (CO$_2$
and H$_2$O) in embedded objects, such as the envelopes of young stars.
Moreover, the 15.2 $\mu$m resonance due to CO$_2$ develops a shoulder
on the red side and gradually shifts towards longer wavelengths
overall, when it is mixed with methanol (CH$_3$OH)
\citep{Ehrenfreund_99_Laboratory}. For instance \citet{Zasowski_09_Spitzer} have
shown that studying the exact shape of the 15.2 micron absorption
feature due to CO$_2$ at a resolution of $R \approx 600$ can indeed be
used as a tool to determine the composition of the ice mixture, and
establish the degree of processing for different
objects. Specifically, they find that in a sample of 16 Class I/II
objects the typical ice composition is dominated by water ice, with
$\sim$12\% CO$_2$, and up to $\sim$10\% of methanol ice. The
ice absorption features at shorter wavelengths (6-12 $\mu$m) also provided
constraints on further ice components.

\begin{figure}[!ht]
\begin{center}
   \resizebox{0.5\hsize}{!}{
     \includegraphics*{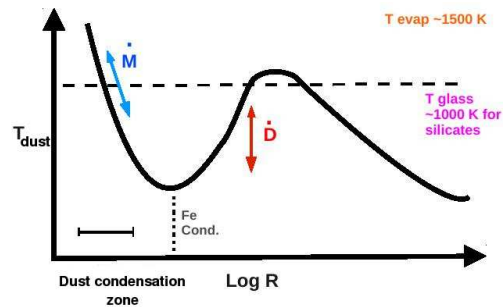}
   }
\end{center}
\caption{Schematic of the dust temperature profile as a function of
  distance to the central star. The dotted line indicates the level of
  the glass temperature, above which silicates can become
  crystalline. Depending on the gas density (defined by the mass-loss
  rate $\dot{M}$), dust condensation starts at a different distance
  from the star, indicated by the blue arrow. This decides whether the
  silicates are formed crystalline or amorphous. As the grains move
  outwards, the temperature drops until it is low enough for iron to
  condense, which cause the UV/optical opacity to increase, and thus
  the grain temperature can be raised above the glass temperature
  (annealing). Figure adapted from
  \citet{Sogawa_99_Origin}.}\label{fig:annealing}
\end{figure}

A second topic of great interest in the wavelength range covered by
MRS is the thermal processing of silicates. Silicates observed in the
interstellar medium are probably not of stoichiometric composition,
and show many lattice defects, often referred to in the
astromineralogy community as being \emph{amorphous} \citep[see
e.g.][]{Kemper_04_Absence}. These amorphous silicates only show broad
resonances due to the Si-O stretching mode at 9.7 $\mu$m, and the
O-Si-O bending mode at 18 $\mu$m. However, when silicates are exposed
to temperatures above the glass temperature ($\sim 1000$ K), annealing
occurs, and the silicates may become crystalline. The spectral
signature of crystalline silicates is quite different, and show a
wealth of narrower resonances due to increased regularity in the
lattice \citep[See Fig.~1 of][]{Molster_05_Crystalline}.  Although it is
known that an increased level of crystallinity is due to exposure of
the silicates to high temperatures, it is not clear whether this is
due to annealing of amorphous silicates, as for instance modeled by
\citet{Sogawa_99_Origin}, or from direct condensation in the gas phase
(See Fig.~\ref{fig:annealing}). Recently, \citet{Jones_12_metallicity} used
a sample of oxygen-rich AGB stars from the Milky Way, the LMC and the
SMC, representing three different metallicities, to show that the
transition between sources showing crystalline silicates or sources
without them, is much sharper for the dust mass-loss rate, than for
the gas mass-loss rate, suggesting that annealing (heating of
amorphous silicates) is the dominant crystallization mechanism.

Finally, a recent exciting discovery is the detection of interstellar
C$_{60}$ \citep{Cami_10_Detection}, by its infrared resonances at 7.0,
8.5, 17.4 and 18.89 $\mu$m. The line strengths in these four bands
provide clues to the excitation and formation mechanisms of this
enigmatic molecule \citep{Bernard-Salas_12_Excitation,Micelotta_12_Formation}, and its chemical
closeness to PAHs may also help understand the formation of these
pre-biotic molecules.

\subsection{Spectroscopy with SAFARI}

With the SAFARI instrument, the 38-55 $\mu$m range is spectrally
accessible for the first time in well over a decade, after the
Infrared Space Observatory (ISO) was able to do this. SAFARI will
operate at 34-210 $\mu$m, with different spectral resolutions, namely
$R\approx 50$, $R\approx 500$ and $R\approx 2000$. Especially
$R\approx 500$ is very suitable for mineralogical studies in
interstellar and circumstellar environments, and here we highlight
three of possible research areas that can be further explored.

First, in 2002, the presence of carbonates in two planetary nebulae was
discovered using the features of dolomite at 62 $\mu$m and calcite at
92 $\mu$m \citep{Kemper_02_Detection}. This was the first extrasolar
detection of these minerals, and their discovery in this
non-parent-body environment suggest a formation path in the absence of
liquid water. An archival search of the spectra of young stellar
objects observed with ISO-LWS revealed a further 17 detections of
carbonates \citep{Chiavassa_05_90}, but additional detections with
\emph{Herschel} have not been reported yet, due to the difficulty of
detecting broad bands in the PACS spectroscopy. With SAFARI a
systematic survey into the occurrence of these resonances can be
performed.

Second, in the SAFARI wavelength range, two water ice features, at 43
and 63 $\mu$m may appear in emission. Previously, these emission
features have been observed towards YSOs \citep{Malfait_99_ISO}
and post-AGB stars \citep{Hoogzaad_02_circumstellar} with ISO-LWS. Water ice is
less volatile than CO$_2$ ice, sticking to dust grains at higher
temperatures, and the ratio between the CO$_2$/H$_2$O mass fractions
informs us about the amount of thermal processing of the
ice. Moreover, the shape and appearance of the 43 and 63 $\mu$m
resonances depend on the lattice structure of the water ice, providing
further information on the formation and processing history.

Finally, SAFARI will offer a further opportunity to study the 69
$\mu$m resonance due to forsterite, initially discovered with
ISO-LWS. This crystalline silicate feature is very sensitive in peak
position to both composition (Fe-content) and dust temperature
\citep{Molster_02_Crystalline_b}, and provides a valuable constraint to dust
models based on crystalline silicate detections at shorter
wavelengths. This feature is narrow enough that it could reliably be
detected with \emph{Herschel}-PACS \citep{Sturm_10_First}, but so
far it has only been analyzed for a handful of objects, allowing for
further discovery space to be filled in by SAFARI observations.

\acknowledgements 
FK wishes to acknowledge support from the National Science Council
under grant number NSC100-2112-M-001-023-MY3.


\end{document}